%% file: REBs_Kepler_BDs_Language.tex
\documentclass{aa}  

\usepackage{graphicx}
\usepackage{array}
\usepackage{graphicx}
\usepackage{lscape}
\usepackage{longtable}
\usepackage{natbib}
\usepackage{color}
\usepackage{array} 
\bibpunct{(}{)}{;}{a}{}{,} 

\usepackage{txfonts}
\begin{document}

   \title{Search for light curve modulations among  \emph{Kepler} candidates}
   \subtitle{Three very low-mass transiting companions}

   \author{J. Lillo-Box\inst{1}\thanks{The first two authors contributed equally to the work contained in this paper.}, A. Ribas\inst{2\star},  D. Barrado\inst{3}, B. Mer\'in\inst{4}, H. Bouy\inst{3} } 

\institute{European Southern Observatory (ESO), Alonso de C\'ordova 3107, Vitacura, Casilla 19001, Santiago de Chile (Chile)\\
              \email{jlillobox@eso.org}  \and
Department of Astronomy, Boston University, 725 Commonwealth Avenue, Boston, MA 02215, USA \and
 Depto. de Astrof\'isica, Centro de Astrobiolog\'ia (CSIC-INTA), ESAC campus 28692 Villanueva de la Ca\~nada (Madrid), Spain\and
European Space Astronomy Centre (ESA), 28691 Villanueva de la Ca\~nada, Spain}
            
  \titlerunning{A search for light curve modulations among  \emph{Kepler} candidates}
\authorrunning{Lillo-Box et al.}
   \date{In prep.}

 
  \abstract
   {Light curve modulations in the sample of \emph{Kepler} planet candidates allows the disentangling of the nature of the transiting object by photometrically measuring its mass. This is possible by detecting the effects of the gravitational pull of the companion (ellipsoidal modulations) and in some cases, the photometric imprints of the Doppler effect when observing in a broad band (Doppler beaming).}     
   {We aim to photometrically unveil the nature of some transiting objects showing clear light curve modulations in the phase-folded \emph{Kepler} light curve. }
   {We selected a subsample among the large crop of  \emph{Kepler} objects of interest (KOIs) based on their chances to show detectable light curve modulations, i.e., close ($a<12~R_{\star}$) and large (in terms of radius, according to their transit signal) candidates. We modeled their phase-folded light curves with consistent equations for the three effects, namely, reflection, ellipsoidal and beaming  (known as REB modulations).}
   {We provide detailed general equations for the fit of the REB modulations for the case of eccentric orbits. These equations are accurate to the photometric precisions achievable by current and forthcoming instruments and space missions. By using this mathematical apparatus, we find three close-in {very low-mass companions (two of them in the brown dwarf mass domain)} orbiting main-sequence stars (KOI-554, KOI-1074, and KOI-3728), and reject the planetary nature of the transiting objects (thus classifying them as false positives). In contrast, the detection of the REB modulations and transit/eclipse signal allows the measurement of their mass and radius that can provide important constraints for modeling their interiors since just a few cases of low-mass eclipsing binaries are known. Additionally, these new systems can help to constrain the similarities in the formation process of the more massive and close-in planets (hot Jupiters), brown dwarfs, and very low-mass companions.}
   {}


   \maketitle
%

\section{Introduction}

In the last two decades, the populations of Jupiter-like planets in close-in orbits around their host stars have been continuously growing. Although the most frequently used detection techniques (i.e., transits and radial velocity) are more efficient for this type of planets, the process by which these hot Jupiters have reached these close orbits is still puzzling. The two competing planet formation mechanisms (i.e., core accretion and gravitational instability) allow the growth of these bodies at large radii (namely $\sim 5-20$~AU for core accretion and $>50$~AU for the gravitational instability model). After the detection of gas giants at orbital separations below 0.5~AU, it was necessary to develop a planet migration theory allowing an inward migration of these planets to closer orbits \citep[see ][and references therein]{chambers09}. The detection of massive planets and substellar objects in close-in orbits around their hosts can provide hints about the actual formation mechanism or at least the necessary conditions in which each of the two mechanisms play a role. 

The low number of well-characterized brown dwarfs (with known radius and mass) prevents a contrast between theory and observations. Different theoretical studies suggest that these substellar objects present large radii during the first stages of their formation while they are contracting. Later on, they can have Jupiter-like radii (or even smaller). However, in  a few cases it was possible to directly quantify the radius of these objects. \cite{stassun06} characterized the radius of a young brown dwarf pair in the Orion Nebula cluster for the first time, confirming a large radius for this object in a $1^{+2}_{-1}$ Myr old star forming region \citep{hillenbrand97,palla99}. Apart from this, to our knowledge, just a few more brown dwarfs have been characterized by deriving accurate radius. Hence, measuring masses and radii of these substellar objects may help  to constrain theoretical models. Additionally, their formation could be similar to that of hot Jupiters and so discovering close-in brown dwarfs in binary systems can help to constrain this scenario.

In this context, the \emph{Kepler} mission has provided an unprecedented photometric precision, delivering long baseline light curves (LC) for more than 150\,000 stars with typical precisions of a few tens of parts per million (ppm) \citep{borucki10, batalha13, burke14}. This impressive precision has mainly been used to detect planetary transits and eclipses where the radius of the transiting companions can be measured. 

Additionally, modulations in the out-of-transit region have been detected in several systems. These ellipsoidal variations are  caused by tidal effects and are used to photometrically measure the mass of transiting objects and to confirm (or reject) their planetary nature \citep[see, e.g., the confirmation of  Kepler-91b by ][]{lillo-box14}.  Ellipsoidal variation is one  of the three major effects detectable in the out-of-transit time interval of the LC, together with the reflection and thermal emission of the planet and the photometric signature of the induced radial velocity, known as Doppler beaming. These three effects (collectively called REBs)  modulate the LC of the star according to the position of the companion along its orbit. In particular, the modulations produced by ellipsoidal variations are characterized by a double bump in the out-of-transit time interval of the LC. Interestingly, the amplitude of these variations depends strongly on the mass of the companion object and the distance from the host star in units of stellar radii (i.e., $a/R_{\star}$). The more massive and the closer  the orbit is, the larger the amplitude of this effect is.

Detecting the transit/eclipse of a companion object and the tidal modulations induced by its closeness to the stellar host allows the simultaneous measurement of its mass and radius. Also, the companion has to be close-in to have any significant effect in the LC. By using \emph{Kepler} data from the prime mission, {we have looked for LC modulations in the sample of \emph{Kepler} objects of interest} (KOIs) with close transiting companions.

 It is also important to note that planets around hot stars (i.e., O- to F-type host stars) are difficult to detect owing to the small number of spectral lines, usually very broad, in their spectra (given their high effective temperatures and rapid rotation of several tens of kilometers per second). Instead, the REB modulations are detectable provided that the companion is close enough and/or massive enough. Therefore, their detection can serve to populate the more massive regime of planet hosts. 
 
 In this paper we present the confirmation of three {very low-mass companions (two within the brown dwarf mass regime)} to main-sequence stars in the sample of \emph{Kepler} planet candidates. In section \S~\ref{sec:sample}, we describe the sample selection, the \emph{Kepler} data, and the processed light curves. In Sect.~\S~\ref{sec:analysis}, we present the general equations for fitting the light curve modulations, including eccentric orbits, and the different implications of the different harmonics. The results are shown in Sect.~\S~\ref{sec:results}, and the implications of these discoveries are discussed in Sect.~\S~\ref{sec:discussion}.

\section{Sample selection and observations \label{sec:observations}}
\label{sec:sample}

\subsection{Sample selection}

REB signatures are stronger for massive close-in objects.  Considering the nominal \emph{Kepler} sensitivity limit of 10 ppm \citep{borucki10}, very low-mass objects cannot create detectable REB signals if $a/R_{\star} > 12$. We therefore restrict our sample to those KOIs with $a/R_{\star} \leq 12$, as determined by the public \emph{Kepler} catalogues \citep{batalha13,burke14}. This yielded a sample of 914 planet candidates.

\subsection{Post-processing of \emph{Kepler} data}

Raw \emph{Kepler} light curves suffer from strong systematics produced by a combination of effects from the spacecraft, detector, and environment. Therefore, they need to be treated prior to any analysis. The \emph{Kepler} archive provides calibrated light curves that are corrected via cotrending and detrending vectors computed with different data from the spacecraft. These curves are extremely useful when  searching for planetary transits, but could alter (or even destroy) other kinds of astrophysical signals. For this reason, we chose to process the lightcurves of the 914 KOIs with a methodology aimed at preserving REB signatures. The processing was as follows:

\begin{enumerate}

\item  For each KOI, we downloaded the available long-cadence light curves via the {\it kplr} python interface.

\item For each individual light curve, we masked the corresponding transits to avoid fitting them, using data from the \emph{Kepler} catalogue (e.g., period, ephemeris).

\item We also masked out the transits of other detected KOIs in the system.

\item The resulting curve was then fit via a cubic spline. The nodes for the fit were chosen by splitting the masked curve in periods.  The phase and flux for each node were then computed as the median values in time and flux, respectively, of the unmasked data within the corresponding bin. By placing nodes with a frequency similar to the planet transit, we make sure that we are neither fitting nor introducing any signal with such a period or smaller. This procedure is standard and has been  used in many other works \citep[e.g., ][]{angerhausen14}.

\item The obtained spline was used to normalize the corresponding long-cadence light curve, including transits.

\item Once every individual light curve was processed, we phased and combined them to create the final curve using the period provided in the \emph{Kepler} catalogue.

\end{enumerate}

After this process, we binned the folded light curves using 500 points across the  phase for visual inspection. We first checked that the obtained light curves were consistent with those of previously studied objects in the literature. In particular, we note that features such as secondary transits or planet reflections are preserved in our processed light curves. 

\subsection{Visual inspection of \emph{Kepler} light curves}

We proceeded to visually inspect each phase-folded light curve, searching for trends resembling REB signatures caused by exoplanets or brown dwarfs. Given the large number of factors that could affect the shape of the final curves, we chose to select only those KOIs with a clear signature (i.e., clearly detectable by eye). The analysis of weak, tentative signals requires much more analysis and is outside of the scope of this paper. Among the several promising candidates, we selected those with no signs of binarity in high-spatial resolution imaging (see Sect.~\ref{sec:highres}):  KOI-554.01, KOI-1074.01, and KOI-3728.01. The remaining possible targets will be modeled in the future when additional data allows a proper estimate of their stellar parameters. We note that we can reproduce the REBs found in the light curve of already confirmed planets by this methodology such as Kepler-91\,b \citep{lillo-box14,lillo-box14c} or KOI-13.01 \citep{mazeh12}.

\subsection{High spatial resolution images: Contaminants in the \emph{Kepler} apertures}
\label{sec:highres}

Owing to the large pixel size of the \emph{Kepler} charge-couple device (CCD) and the large apertures used by the \emph{Kepler} team to extract the precise photometry (6-10 arcsec), possible close sources can contaminate the light curves of the preselected objects. Several teams have performed ground-based follow-up of the planet candidates by obtaining high-spatial resolution images of the host stars to detect these possible contaminants and be able to correct the light curves. Several techniques have been applied such as adaptive optics in the near-infrared \citep[e.g.,][]{adams12,adams13}, adaptive optics in the optical range \citep[e.g.,][]{law13}, speckle imaging in the optical \citep[e.g.,][]{howell11}, or lucky-imaging in the optical \citep[e.g.,][]{lillo-box12,lillo-box14b}. In the last work, we performed a comparison between the different surveys and techniques by means of the newly defined  blended source confidence (BSC) parameter. This parameter provides the probability that a given source detected as being isolated in the high-spatial resolution image has no undetected companions within a certain magnitude range and angular separation\footnote{ The calculation of the BSC will be made available in python- and IDL-based codes, as defined in \cite{lillo-box14b}.}. 

Lucky-imaging observations of KOI-3728 with AstraLux (Calar Alto Observatory) were analyzed in \cite{lillo-box14b}. For KOI-554 and KOI-1074, high spatial resolution images were retrieved from the Community Follow-up Observing Program\footnote{https://cfop.ipac.caltech.edu/home/} (CFOP) of the \emph{Kepler} mission and were obtained and analyzed by David Ciardi. We only found a companions closer than 6 arcsec in the case of KOI-554. However, by overlaying the \emph{Kepler} apertures on the high-resolution image, we can see that this companion is always out of the apertures used in Q1-Q17, so that it does not affect the photometry of the prime target. In Table~\ref{tab:astralux} we show the analysis results of the high spatial resolution images. In the case of KOI-3728,   observed by our team with the AstraLux instrument, we obtain a BSC = 99.6\%. According to these observations we can assume hereafter that the KOIs are isolated.

\input{Table_AstraLuxResults.tex}

\section{Light curve modeling \label{sec:analysis}}

We model the full light curves, which are composed of  the contribution from the transit/eclipse and the out-of-transit modulations induced by massive companions. Here, we describe the adopted procedure.

\subsection{Eclipse signal}

The eclipses were modeled by using the \cite{mandel02} prescriptions. We assumed the stellar properties ($T_{\rm eff}$, $\log{g}$, and [Fe/H]) -- publicly available on the CFOP of the \emph{Kepler} mission and published in \cite{huber14} -- to determine the limb darkening and gravity darkening coefficients (see Table~\ref{tab:ld}). These coefficients were obtained by trilinearly interpolating the \cite{claret11} tabulated values on the estimated stellar properties. We performed a bootstrapping on the stellar properties and their corresponding uncertainties to measure their impact on the limb darkening coefficients. The resulting relative errors of these parameters are around 2-5\%, which is much smaller than the inferred  uncertainties of the parameters. This given, the transit/eclipse model depends on six free parameters: planet-to-star radius ($R_p$), orbital eccentricity ($e$), argument of the periastron ($\omega$), semi-major axis ($a$), stellar radius ($R_{\star}$), and orbital inclination from the plane of the sky perpendicular to our line of sight ($i$).

\input{Table_LimbDarkening.tex}

\subsection{Light curve modulations}

In this section we present a complete set of the equations used to model these photometric effects for any type of orbital configuration (including eccentric orbits).   \\

\noindent \textbf{Planet reflection and emission. } 
The reflected light by the companion is described by the inverse ratio of the orbital distance and its radius ($R_p/d$), the ratio between the incident and reflected light (also known as the albedo, $A_g$), and the phase function of the companion generally expressed as $\Phi(z)$. The last function is formulated in two main recipes, called the Lambertian reflection phase function, $\Phi_{\rm Lam}(z)$, and the geometric phase function, $\Phi_{\rm geo}(z)$. As demonstrated by \cite{faigler11}, $\Phi_{\rm Lam}(z) = \Phi_{\rm geo}(z) + 0.18\sin^2i ~\cos{(2\theta)} + C$, where the term C encompasses other constant and negligible terms. Determining which of the two phase functions is more correct is not possible with just the \emph{Kepler} data. By assuming one or the other, we can introduce a double-peak signal that can increase the ellipsoidal amplitude by up to 18\%, artificially increasing or decreasing the mass of the companion if we assume the incorrect phase function. Regardless of the assumed phase function, the expression for the reflected light of the companion is given by

\begin{equation}
\label{eq:reflection}
\left( \frac{\Delta F}{F} \right)_{\rm ref} = A_g \left(\frac{R_p}{d}\right)^2 \Phi(z)  ~ \equiv ~ A_{\rm ref} \Phi(z)
\end{equation}

Throughout this work we  assume the Lambert approach to ensure that the obtained masses are the highest of the two possibilities. \\

\noindent \textbf{Ellipsoidal variations. }
The ellipsoidal variations are produced by the gravitational pull of the close companion on the host star. This effect was theoretically investigated by \cite{pfahl08}, who determined the flux variations due to stellar oscillations induced by close substellar companions. Based on this theoretical analysis, the ellipsoidal modulations can be expressed as a sum of different harmonics. In particular, $l=2$ and $l=3$ are the only relevant values in this work. Indeed, the third harmonic ($l=3$) has only been detected 
  in the well-known case of KOI-13.01 \citep{mazeh12,esteves13}. Higher order harmonics ($l\geq4$) induce amplitudes more than one order of magnitude smaller than $l=2$ for $a/R_{\star} \gtrsim 3$. Only for extremely close-in companions ($a/R_{\star} \lesssim 2$) will higher harmonics  play a relevant role for the current and forthcoming achievable photometric precisions. Then, we can neglect harmonics with $l\geq 4$, and so the relative flux variations caused by a companion of mass $M_p$ at an orbital separation of $d/R_{\star}$ in an orbit with an inclination $i$ around a star of mass $M_{\star}$is given by
\begin{equation}
\label{eq:ellip_all}
\begin{split}
\left(\frac{\Delta F}{F} \right)_{\rm ellip} =  ~ \frac{M_p}{M_{\star}} ~ \left( \frac{R_{\star}}{d} \right)^3 ~   \overbrace{ \left(  \frac{1}{4} f_2\left[ -(3\cos^2{i}-1) + 3\sin^2{i}\cos{2\theta} \right] \right.}^{l=2} + \\
   + \underbrace{ \left. \frac{1}{8}\left( \frac{R_{\star}}{d} \right) f_3 \sin{i} \left[ -3(5\cos^2{i}-1) \cos{\theta} + 5\sin^2{i}\ \cos{3\theta} \right]  \right) }_{l=3},
\end{split}
\end{equation}

\noindent where the $f_1$ and $f_2$ coefficients are defined in \cite{pfahl08}. These coefficients can be written in terms of the linear limb-darkening and gravity darkening coefficients. From Eqs. 9-13 in \cite{esteves13} we can identify that $f_2 = 4\alpha_2/3$ and $f_3 = 8\alpha_1 \alpha_2$, being
\begin{equation}
\label{eq:coefficient}
\begin{split}
\alpha_1  = & \frac{25u}{24(15+u)} \left(\frac{y+2}{y+1} \right) \\ 
\alpha_2  = & \frac{3(15+u)}{20(3+u)} \left(y+1 \right). \\ 
\end{split}
\end{equation}

In cases where the third harmonic is also negligible (e.g., for highly inclined orbits or large orbital separations), the relative variations can be re-written as 

\begin{equation}
\label{eq:ellip_2nd}
\begin{split}
\left( \frac{\Delta F}{F} \right)_{\rm ellip} \approx ~ & \left[ \frac{3f_2}{4}  ~ \frac{M_p}{M_{\star}} ~ \left( \frac{R_{\star}}{d} \right)^3 ~  \sin^2{i}\   \cos{2\theta} \right]  \\
                       &   - \left[ \frac{M_p}{M_{\star}} ~ \left( \frac{R_{\star}}{d} \right)^3 ~ \frac{f_2}{8}  (3\cos^2{i}-1)  \right].
\end{split}
\end{equation}

We note that the orbital separation between the two objects is given by $d/R_{\star} = a/R_{\star} (1-e^2)/(1+e\sin{\omega})$. Thus, in the particular case of near-circular orbits where $d\approx a = $~constant, the second term of the right-hand side of this equation is constant and can then be neglected. This yields the commonly used equation for this type of modulations with the $\cos{2\theta}$ dependency that provides the characteristic double-peak in the phase-folded light curve

\begin{equation}
\label{eq:ellip_simple}
\left(  \frac{\Delta F}{F} \right)_{\rm ellip} \approx \frac{3f_2}{4} ~ \frac{M_p}{M_{\star}} ~ \left( \frac{R_{\star}}{d} \right)^3 ~  \sin^2{i}\   \cos{2\theta} ~\equiv~ A_{\rm ellip}~ \cos{2\theta}.
\end{equation}

In this work, we have used the more complete Eq.~\ref{eq:ellip_all} in our modeling since the same number of free parameters is needed and always accounts for the two relevant harmonics.\\

\noindent \textbf{Doppler beaming. }
The Doppler beaming is the photometric imprint of the radial velocity when we obtain precise photometry at a particular (fixed) bandpass. \cite{loeb03} and more recently \cite{bloemen11} have provided a theoretical expression for this effect, being

\begin{equation}
\label{eq:beaming_complete}
 \left( \frac{\Delta F}{F} \right)_{\rm beam} =  B ~\frac{K}{c}  \sin{\theta},
\end{equation} 

where $K$ is the radial velocity semi-amplitude, c is the speed of light, and $B$ is the beaming factor defined as $B=5+d{\rm Ln}F_{\lambda}/d {\rm Ln}\lambda$, where $F_{\lambda}$ is the stellar flux. In the case of the \emph{Kepler} observations, this factor can be computed as

\begin{equation}
\label{eq:beaming_factor}
<B> = \frac{\int \epsilon_{\lambda} \lambda F_{\lambda} B d\lambda  }{\int \epsilon_{\lambda} \lambda F_{\lambda}  d\lambda},
\end{equation} 

where $\epsilon_{\lambda}$ is the response function of the \emph{Kepler} bandpass\footnote{\url{http://keplergo.arc.nasa.gov/CalibrationResponse.shtml}}.  The stellar flux $F_{\lambda}$ is obtained from the {\sc nextgen} models \citep{hauschildt99} according to the physical parameters of the star shown in Table~\ref{tab:ancillary}.


\section{Results  \label{sec:results}}

We used a Bayesian approach to derive robust uncertainties for the parameters in our models. For this purpose, we used the \emph{emcee} implementation \citep{emcee} of the affine-invariant ensemble sampler for Markov chain Monte Carlo \citep[MCMC; ][]{Goodman2010}.

We then modeled the processed light curves by adding the REB signals to the transits \citep[computed using \emph{PyTransit};][]{pytransit} in the phased curves. Our model contains ten free parameters: mass and radius of the planet ($M_p$, $R_p$), mass and radius of the star ($M_s$, $R_s$), semi-major axis ($a$), orbital inclination ($i$), eccentricity ($e$), argument of the periastron ($\omega$), the beaming factor ($B$), the albedo, and a scaling factor. This last parameter accounts for the fact that the out-of-transit flux is not constant for light curves with REB signals and depends on different factors (shape of REB signal, processing of light curves, etc.). For some of these parameters, we have previous information from the CFOP or they were previously calculated by using the CFOP  stellar parameters (from \citealt{huber14}) such as the beaming factor. The used mean values and uncertainties are shown in Table~\ref{tab:ancillary}. These values are included with Gaussian priors, which are listed in Table~\ref{tab:priors}.

\input{Table_Ancillary.tex}

\input{Table_Priors.tex}

With this setup, we ran 20000 iterations with \emph{emcee} using 50 walkers. We discarded the first 15000 iterations to ensure that convergence had been reached in all cases, which was confirmed via visual inspection of the chain evolution. The remaining chain positions trace the posterior distributions, and can be used to estimate different confidence intervals for each parameter. The resulting median value and the 15.9-84.1\,\% confidence intervals (corresponding to $1\sigma$ in Gaussian distributions) for each target are listed in Table~\ref{tab:priors}. We find three of the KOIs studied to be low-mass eclipsing binaries, two of which are compatible with being brown dwarfs within 1$\sigma$ uncertainties\footnote{Indeed, KOI-554.01 was classified as a brown dwarf by \cite{santerne12}, but no determination of its mass is found in the literature to our knowledge.}.

{In the case of KOI-1074.01, the large albedo indicates an intrinsic luminosity of the companion. However, since no secondary eclipse is found, the contamination introduced by this companion is already taken into account in the scaling factor. This is clear when comparing the weighted goodness of fit (Bayesian information criterion, BIC) of the two models: the one presented above and a second model including a dilution factor to account for the brightness of the secondary star and restricting the albedo to be $A_B<1$. The result shows a difference in the BIC value of 12 in favor of the former (simpler) model. This is expected since we do not have a significant detection of the secondary. Additionally, since the detection of the REB modulations is less clear in this system, we tested the hypothesis of non-detection by just fitting the transit signal. In this case, we find a difference in the BIC value of $\Delta BIC =63$ in favor of the complex model (including REBs), showing  significant evidence against the non-detection hypothesis. }

\section{Discussion and conclusions\label{sec:discussion}}

In this paper we have presented detailed general equations to model the light curve modulations produced by close-by companions. By making use of the \emph{Kepler} data, we have analyzed three cases where very low-mass companions were found in the sample of planet candidates: KOI-554.01, KOI-1074.01, and KOI-3728.01. These candidates are thus false positive planets and should be catalogued as binary systems. 

\begin{figure}
      \includegraphics[width=0.50\textwidth]{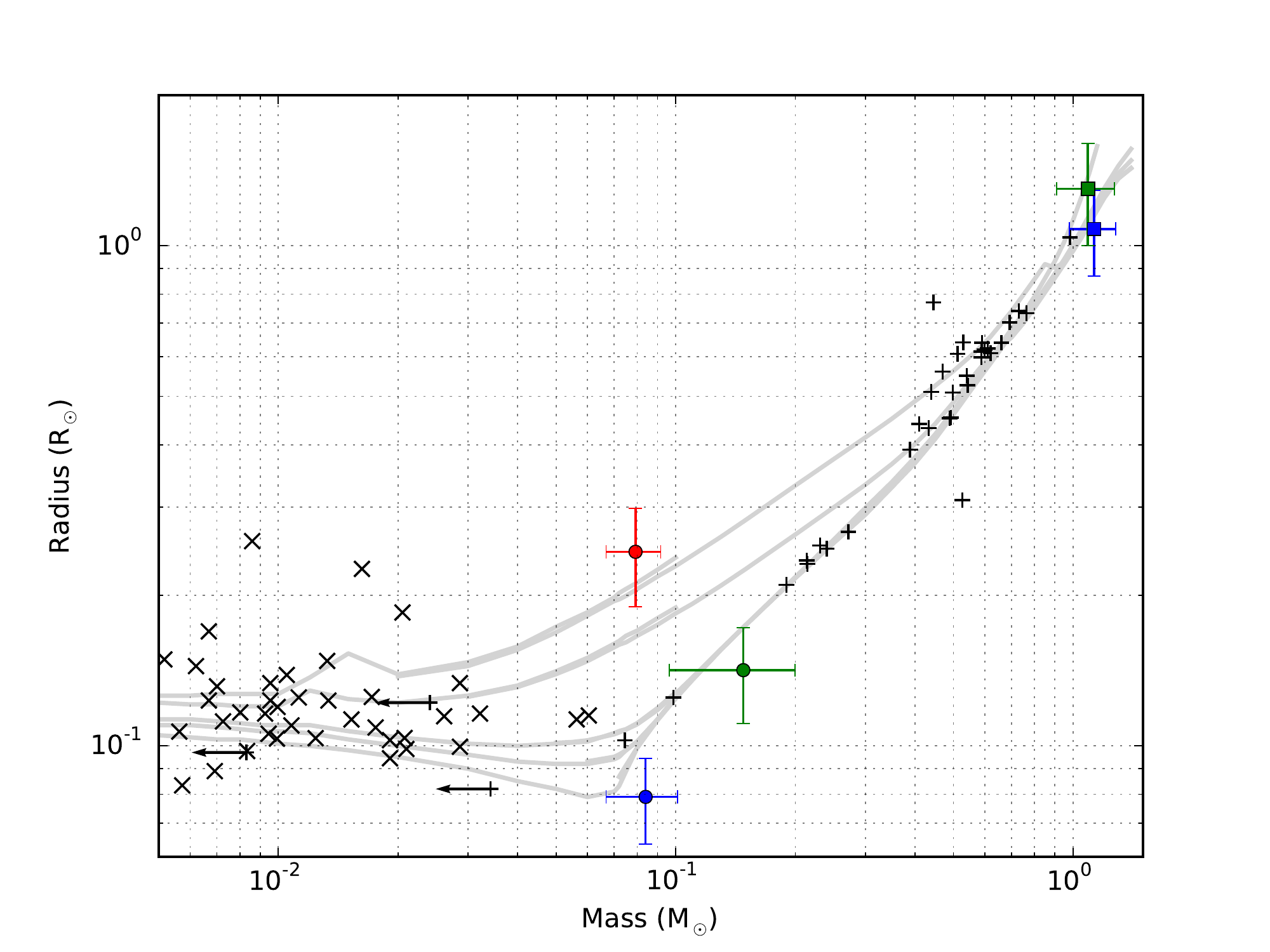}
      \caption{Mass-radius relationship  for the three detected binaries in the \emph{Kepler} sample, namely KOI-554 (blue), KOI-1074 (green), and KOI-3728 (red). Primary stars are represented by squares, while the close companions are represented by circles. Other low-mass eclipsing binaries from the literature are also plotted with black plus symbols \citep{huelamo09, shkolnik08, diaz14, lillo-box15a}. Massive planets from the Exoplanet Catalogue (exoplanet.eu) are plotted as black crosses. Leftward arrows show the upper limits to the masses of the three fast rotators in \cite{lillo-box15a}. Solar metallicity isochrones of \cite{baraffe98} are also plotted for reference as gray thick lines. From top to bottom: 50 Myr, 100 Myr, 500 Myr, 1 Gyr, and 5Gyr. }
   \label{fig:MR}
   \end{figure}

Since they are also transiting, we can measure their basic parameters, namely their mass and their radius.  This has been possible for a few   very low-mass objects \citep[e.g., ][]{huelamo09, shkolnik08, diaz14, lillo-box15a} so these detections increase the sample. This is important in order to understand the properties of these sources since they allow a direct measurement of their density and surface gravity. In Fig.~\ref{fig:MR}, we show the location of the three characterized very low-mass companions together with the known sample of low-mass eclipsing binaries and the more massive confirmed planets from the literature. The three systems studied here (together with a few others) fill a void region in the diagram, part of the so-called brown-dwarf desert. The derived properties of these low-mass companions can feed formation theories and interior models \citep{baraffe14} of objects of this type, since very few of them are known. 

In the particular case of KOI-3728, in \cite{lillo-box15a} we provided a radial velocity based upper limit to the mass of the transiting object of $M_2 \lesssim 116~M_{\rm Jup}$. Here, we found a mass of $M_2 = 83\pm 13~M_{\rm Jup}$, in good agreement with this upper limit. This object orbits a massive star ($M_{\star} = 2.05\pm0.17~M_{\odot}$) in the subgiant phase. It shows a larger radius than expected for the age of the system, which may indicate some inflation. This system is thus in a key step of the evolutionary process when several processes take place (dynamical instabilities, tidal forces, etc.)   and can affect the orbital and physical structure of the components of the system. It has been theorized that the inflation of low-mass short-period binaries \citep[discussed in ][]{ribas06} is produced by the inhibition of convection due to magnetic fields \citep[e.g., ][]{david15}, which could be reinforced by the tidal forces induced during the subgiant phase. Although this is just one possibility, this system can be a test case for these theories. 

Additionally, by constraining the detection of the secondary eclipses (considering them as upper limits to their depths) and comparing their depth with that of the primary transit, we can obtain the ratio of temperatures between the two objects. So, from Table~\ref{tab:ancillary} we can infer an upper limit for the effective temperature of the companion since $\delta_1/\delta_2 = (T_A/T_B)^4$, where A is the primary star and B is the companion, and $\delta_1$ and $\delta_2$ represent the depth of the primary and secondary eclipses, respectively. By doing so, we obtained $T_{\rm 554.01} < 2500$~K, $T_{\rm 1074.01} < 1500$~K, and $T_{\rm 3728.01} < 2100$~K, indicating that the companions are ultracool dwarf companions.\\

We have also provided comprehensive and general equations for light curve modulation analysis. Given the high-precision photometric missions that will be launched in the near future (e.g., \emph{Cheops}, \emph{TESS,} and \emph{PLATO}), the analysis of light curve modulations can provide important results by unveiling the nature of transiting companions. Also, since there is no need for the companion to transit, new planets and low-mass companions can be detected from these surveys. Even though different works have published the equations of the different components of the REB modulations \citep[e.g., ][]{faigler12, faigler14, esteves14, mazeh12}, there is still a discrepancy with the masses measured by radial velocity technique. Understanding these differences is crucial in order to correctly characterize the new detections without the need for a subsequent radial velocity follow-up. Hence more detections are needed to this end. In this work we present three new detections that will help in this task.


\begin{acknowledgements}
      J.L-B acknowledges financial support from the Marie Curie Actions of the European Commission (FP7-COFUND) and the Spanish grant AYA2012- 38897-C02-01. We  appreciate the extraordinary effort and generosity by all the people involved in the {\it Kepler} mission at some point, those running the observatory and  those providing an amazing database with more than a hundred  thousand  light curves and thousands of planetary candidates, so the community can work on them, and can confirm and  characterize the largest and most comprehensive planetary sample found so far, as well as other interesting astronomical systems. This work used the python packages {\it numpy}, {\it matplotlib}, {\it astroML}, {\it asciitable}, {\it pyTransit}, {\it kplr}, and {\it emcee}.
\end{acknowledgements}

\bibliographystyle{aa} 
\bibliography{../biblio2} 

\newpage


   \begin{figure*}
   \includegraphics[width=0.5\textwidth]{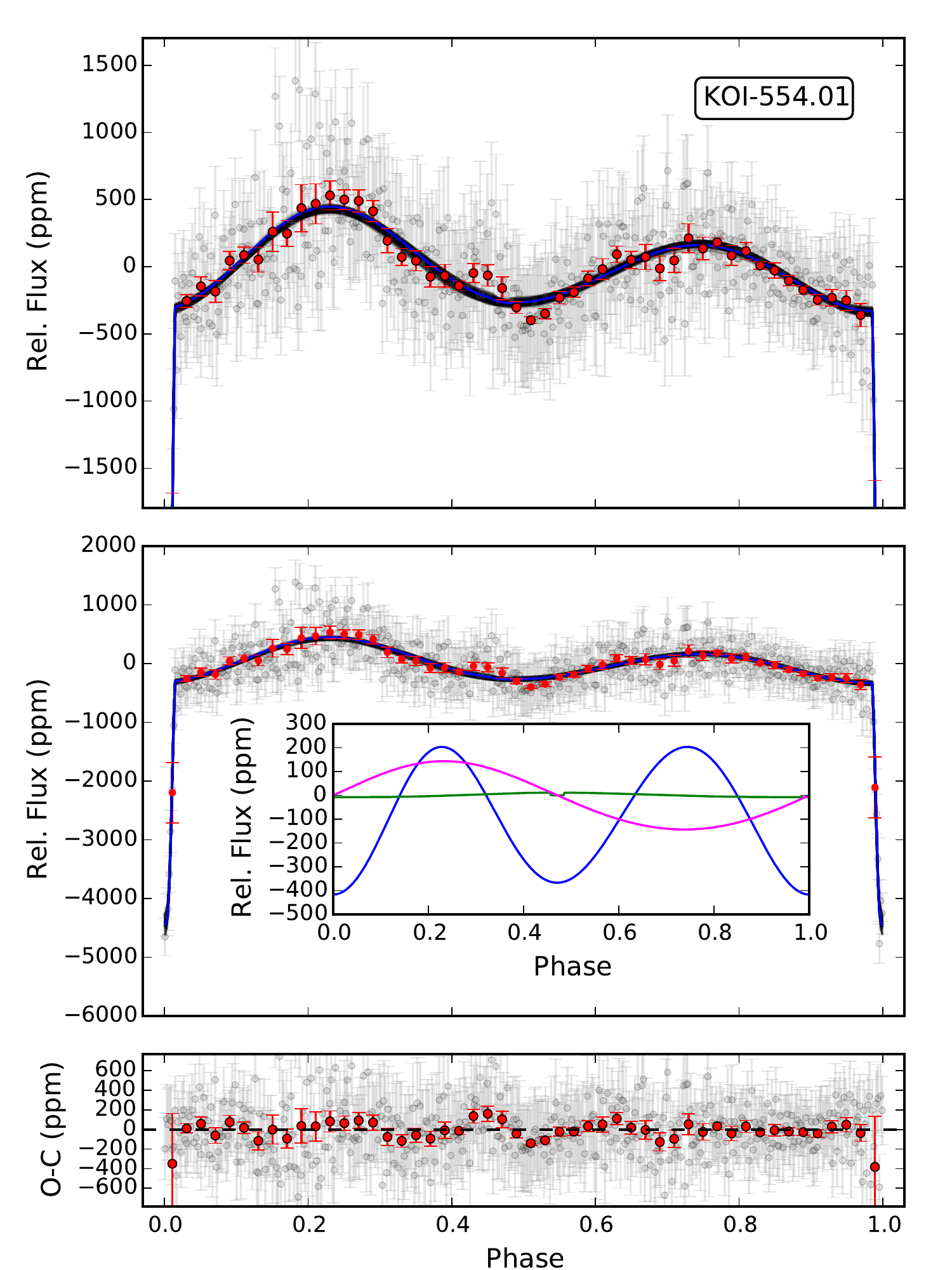}\includegraphics[width=0.5\textwidth]{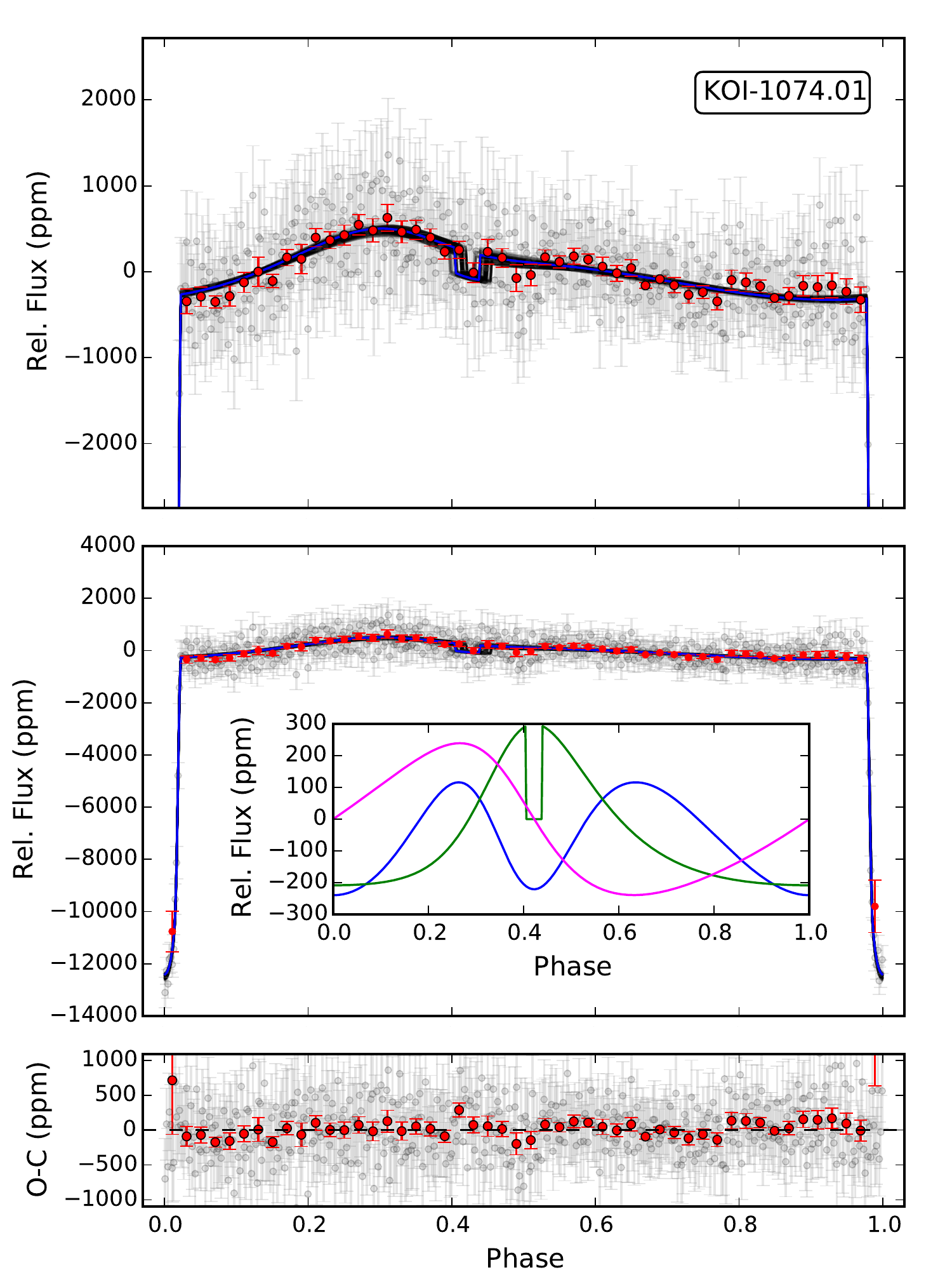}
   
  \includegraphics[width=0.5\textwidth]{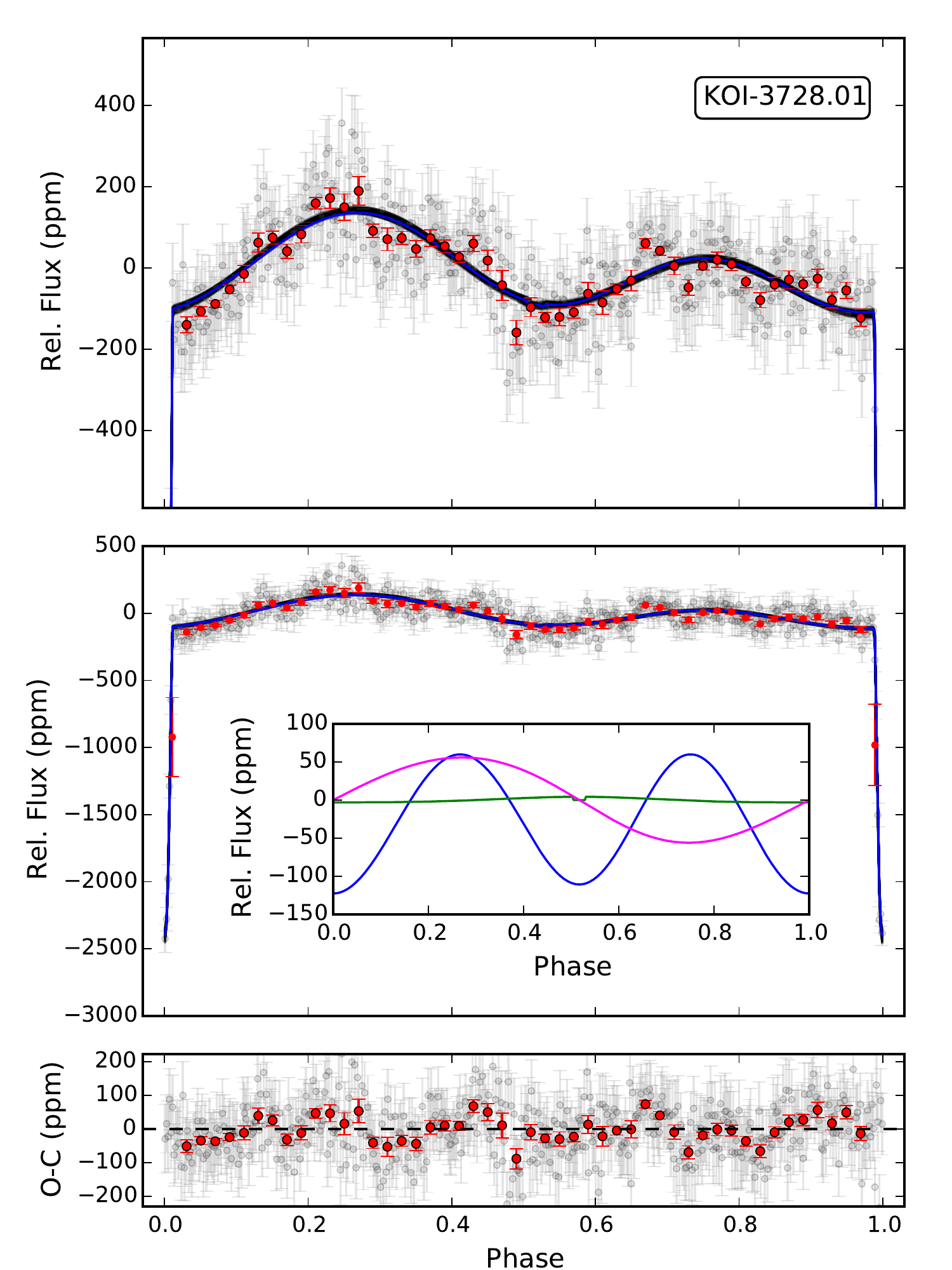}
     \caption{Fitting results for the three KOIs analyzed in this work. For each object, the top panel shows a close view of the out-of-eclipse time interval where the light curve modulations are detectable. The middle panel shows the complex light curve including the eclipse and an inset showing each contribution to the out-of-eclipse modulations. In the bottom panel, we show the residuals of the fit. Gray circles represent 500 bins along the orbit and red circles represent  binnings of  100 datapoints. The best one hundred models are shown with black lines and the best model is shown with a blue line.}
   \label{fig:FittingResults}
   \end{figure*}

   \begin{figure*}
   \includegraphics[width=0.2\textwidth]{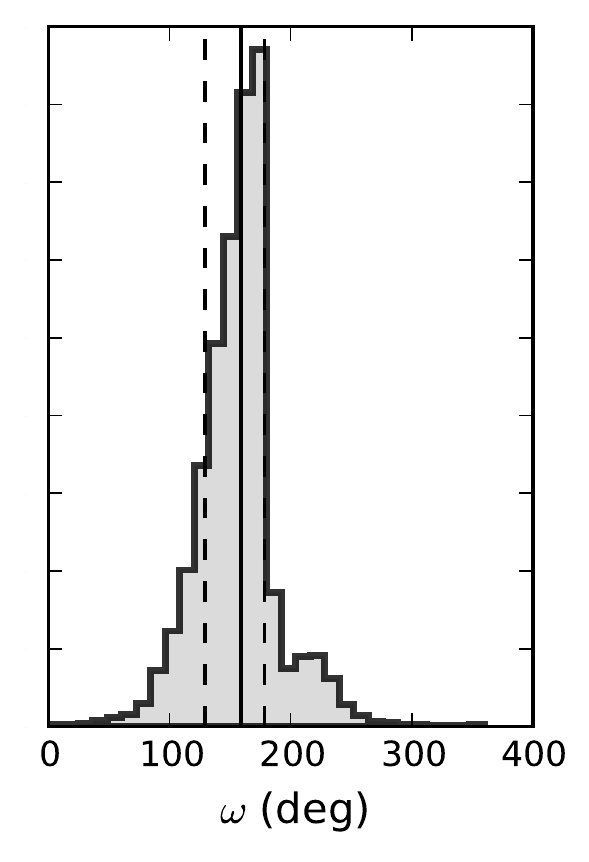}   
   \includegraphics[width=0.2\textwidth]{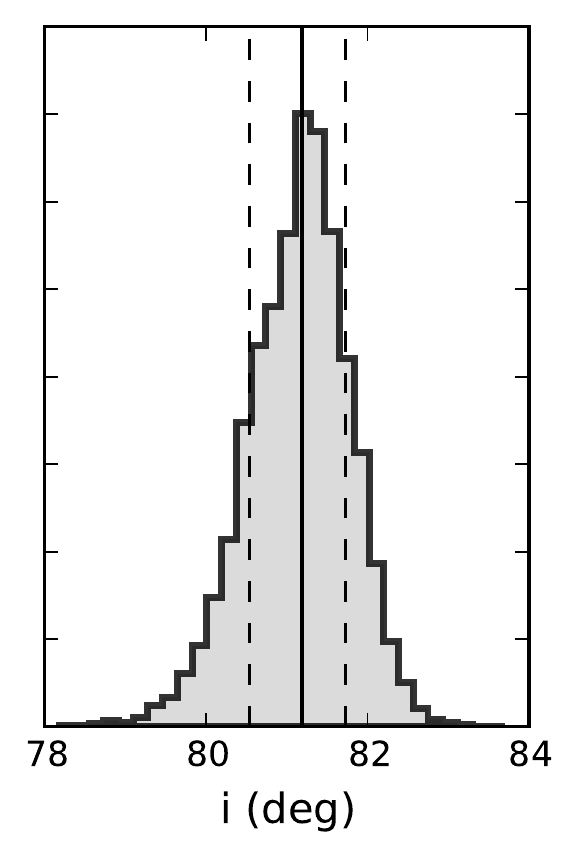}   
   \includegraphics[width=0.2\textwidth]{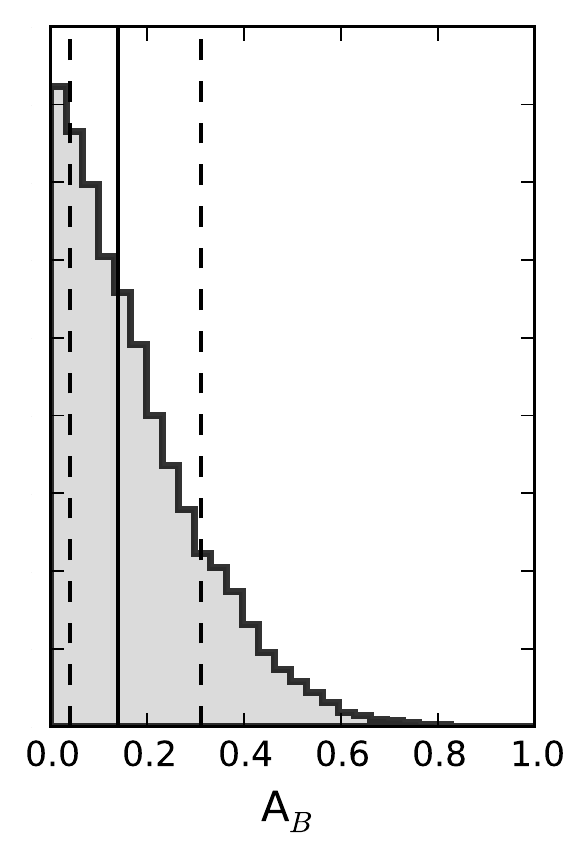}   
      \caption{{Asymmetric} posterior distribution in the analysis of the light curve modulations of KOI-554.01 obtained from an MCMC analysis with 20000 steps with 50 walkers (see Sect.~\ref{sec:results}). Only the last 5000 steps are used to compute the final parameters and to build these posterior distributions. The vertical solid line represents the median value of the distribution, while the dashed lines show the 16.7 and 83.4 percentiles of the distribution.}
   \label{fig:FittingResults}
   \end{figure*}

   \begin{figure*}
   \includegraphics[width=0.19\textwidth]{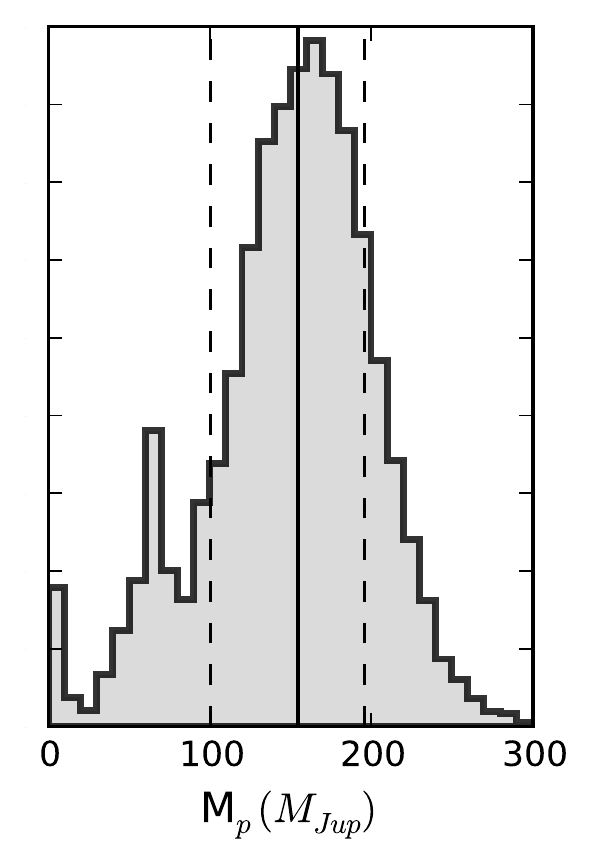}   
   \includegraphics[width=0.19\textwidth]{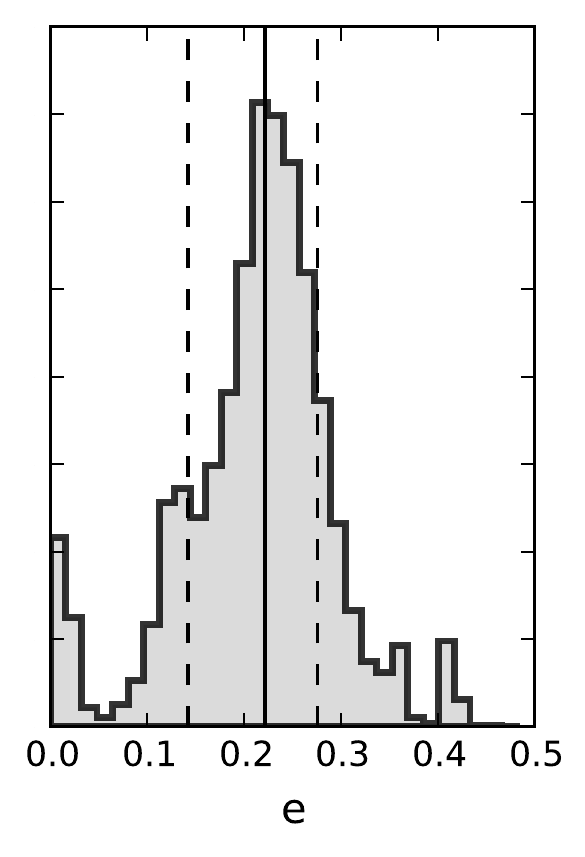}   
   \includegraphics[width=0.19\textwidth]{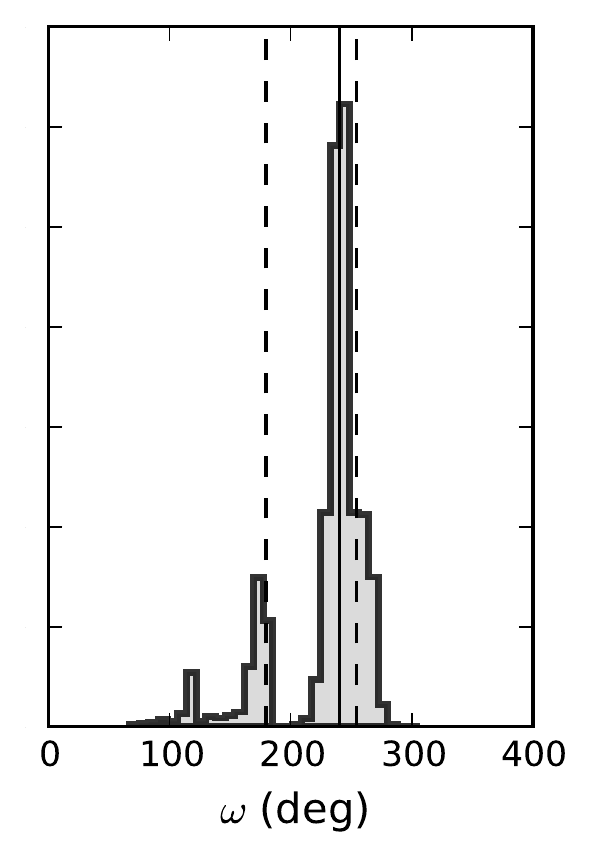}   
   \includegraphics[width=0.19\textwidth]{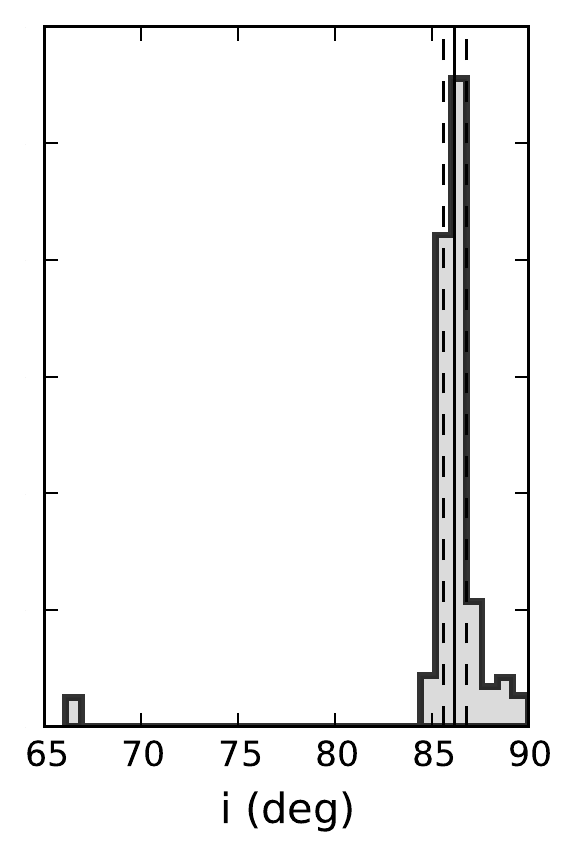}   
   \includegraphics[width=0.19\textwidth]{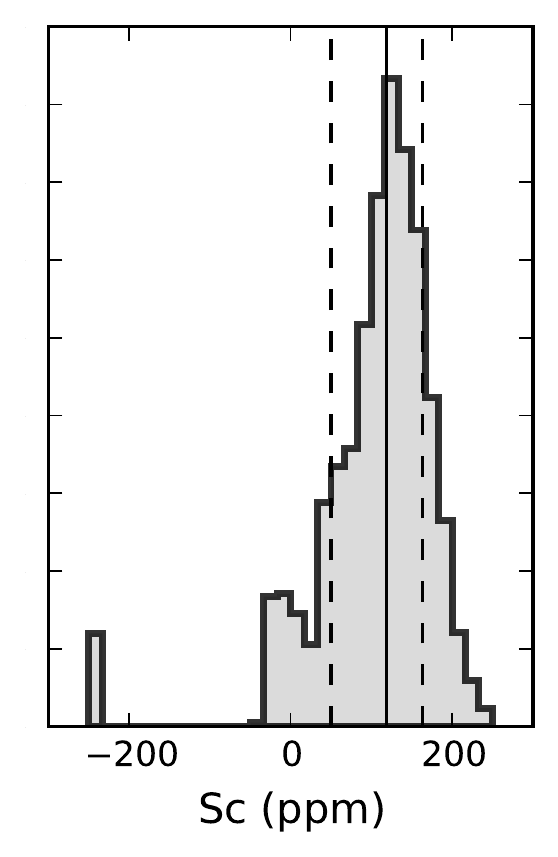}   
      \caption{{Asymmetric} posterior distribution in the analysis of the light curve modulations of KOI-1074.01 obtained from an MCMC analysis with 20000 steps with 50 walkers (see Sect.~\ref{sec:results}). Only the last 5000 steps are used to compute the final parameters and to build these posterior distributions. The vertical solid line represents the median value of the distribution,  while the dashed lines show the 16.7 and 83.4 percentiles of the distribution.}
   \label{fig:FittingResults}
   \end{figure*}

   \begin{figure*}
   \includegraphics[width=0.19\textwidth]{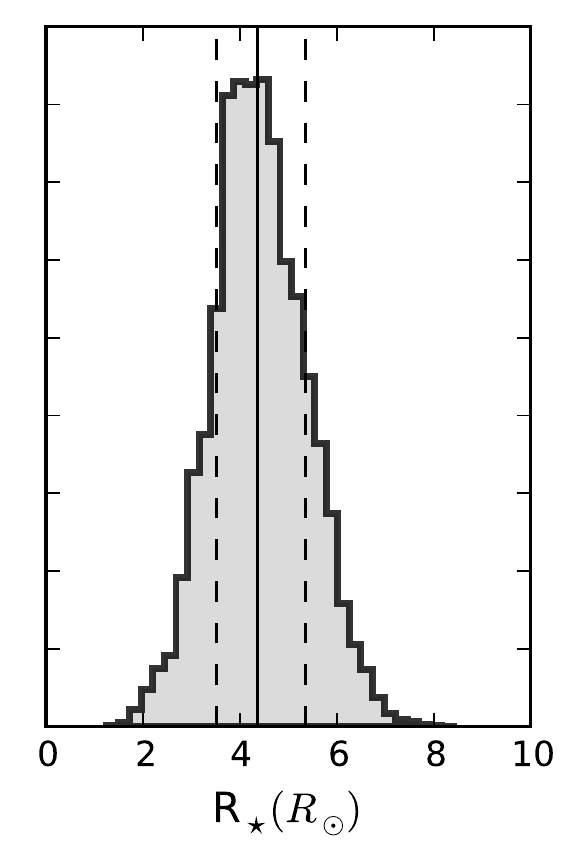}   
   \includegraphics[width=0.19\textwidth]{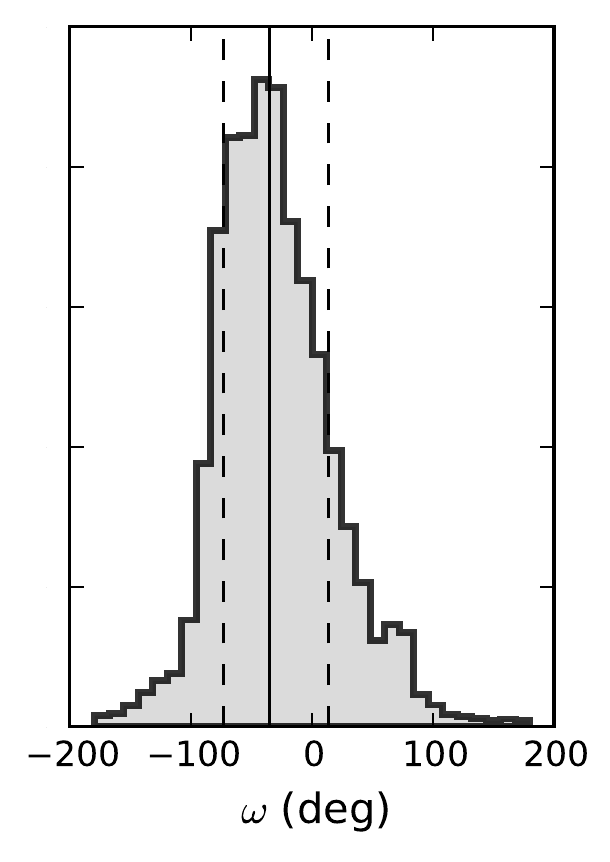}   
   \includegraphics[width=0.19\textwidth]{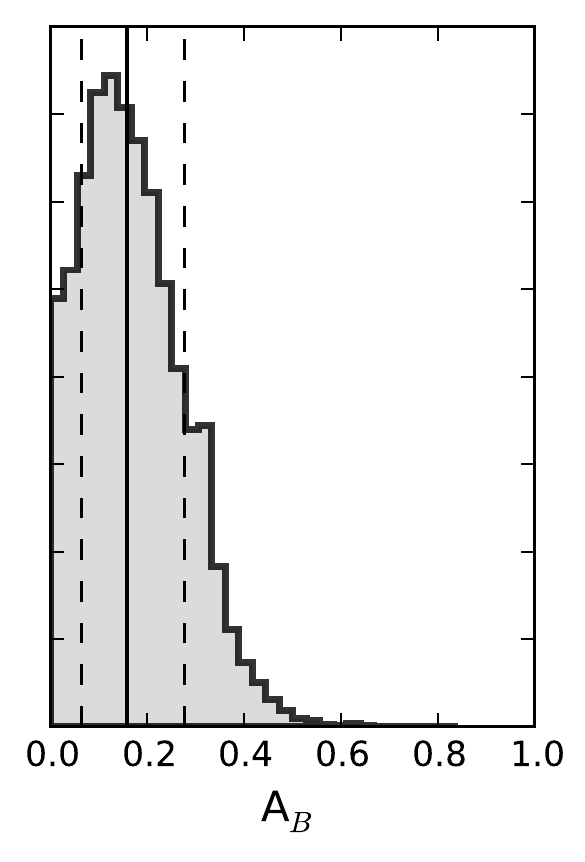}   
   \includegraphics[width=0.19\textwidth]{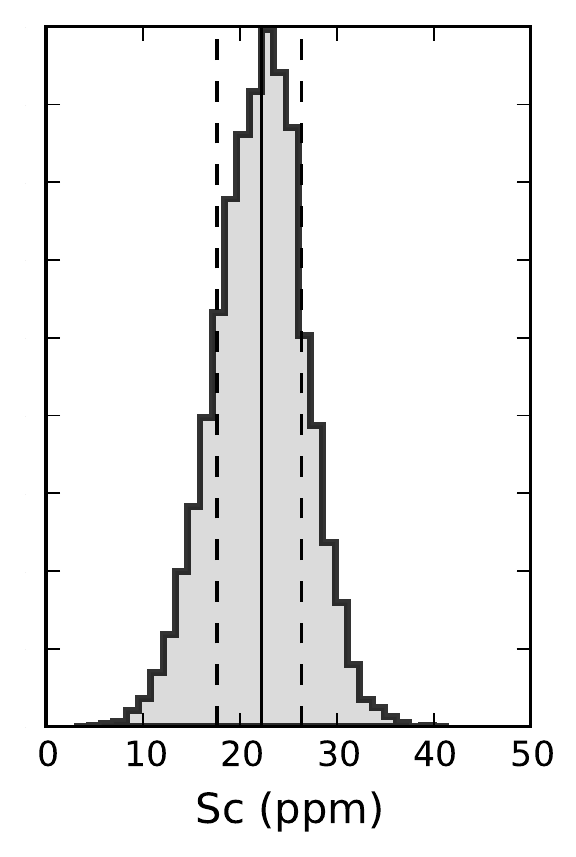}   
      \caption{{Asymmetric} posterior distribution in the analysis of the light curve modulations of KOI-3728.01 obtained from an MCMC analysis with 20000 steps with 50 walkers (see Sect.~\ref{sec:results}). Only the last 5000 steps are used to compute the final parameters and to build these posterior distributions. The vertical solid line represents the median value of the distribution,  while the dashed lines show the 16.7 and 83.4 percentiles of the distribution.}
   \label{fig:FittingResults}
   \end{figure*}

\end{document}

%% file: Table_AstraLuxResults.tex
\begin{table}
\setlength{\extrarowheight}{2pt}
\small
\caption{Ancillary data and calculated  blended source confidence  values for the three KOIs studied in this paper.}             
\label{tab:astralux}      
\centering          
\begin{tabular}{r c c c c c c c}    
\hline\hline       

 KOI      &     High-res im.\tablefootmark{a}  & Comp.  & Sep.  & $\Delta m_{Kep}$   & BSC\tablefootmark{b}         \\
          &     Ref.                           & ($<6$\arcsec) &  arcsec.  & mag & (\%)         \\ \hline 
 554   &    CFOP   & B & 5.65\arcsec & 1.68      & - \\
1074  &    CFOP   & 0 & -      & -       & - \\     
3728  &     LB14   & 0 & -      & -         & 99.6\%  \\     

\hline                  
\end{tabular}
\tablefoot{
\tablefoottext{a}{LB14 = \cite{lillo-box14}, CFOP = no reference found in the bibliography, but companion tables of the UKIRT J-band survey by D. Ciardi are available in the CFOP.}
\tablefoottext{b}{The blended source confidence parameter (BSC) provides the probability that no chance-aligned eclipsing binaries are  responsible for the detected transit signal given a high spatial resolution image while taking into account its sensitivity limits \citep[see ][]{lillo-box14}.}
}

\end{table}

%% file: Table_LimbDarkening.tex
\begin{table}
\setlength{\extrarowheight}{2pt}
\small
\caption{Derived gravity darkening ($y$) and linear ($u$) and quadratic ($q_1$ and $q_2$) limb darkening coefficients based on the stellar properties presented in Table~\ref{tab:ancillary}.}             
\label{tab:ld}      
\centering          
\begin{tabular}{r c c c c}     
\hline\hline       

 KOI      &  $u$   & $y$    & $q_1$  & $q_2$ \\ \hline
 554   & 0.558  & 0.312  & 0.338  & 0.295 \\
1074   & 0.540  & 0.284  & 0.314  & 0.304 \\
3728   & 0.490  & 0.150  & 0.256  & 0.318 \\ 

\hline                  
\end{tabular}

\end{table}

%% file: Table_Ancillary.tex
\begin{table*}
        \setlength{\extrarowheight}{2pt}
        \small
        \caption{Ancillary parameters from the CFOP and determined beaming factor (from Eq.~\ref{eq:beaming_factor}) 
                        for the three KOIs studied in this paper.}             
        \label{tab:ancillary}      
        \centering          
        \begin{tabular}{r c c c c c c}  
        \hline\hline       

        KOI      &   Period &  $M_{\star}$  & $R_{\star}$ & $T_{\rm eff}$ & $\log{g}$  &B    \\
                &     (days) &  ($M_{\odot}$)                              & ($R_{\odot}$)                             & (K)                    &   (cgs) & -    \\ \hline 
        554.01   &  $3.65849447\pm0.00000125$ & $1.07\pm0.14$    & $1.00\pm0.09$ & $6108\pm140$ & $4.64\pm0.05$ & $3.69\pm0.18$ \\  
        1074.01   &  $3.770552098\pm0.000000942$ & $1.07\pm 0.11 $    & $1.10\pm0.12$& $6302\pm180$ & $4.49\pm0.07$ &$3.56\pm0.18$ \\     
        3728.01  &   $5.546083665\pm0.000007087$ & $2.05\pm0.17$      &$4.04\pm0.45$ & $7360 \pm 220$ &$3.54 \pm0.32$  &  $3.14\pm0.15$ \\     

        \hline                  
        \end{tabular}
        \end{table*}

%% file: Table_Priors.tex
\begin{table*}
\setlength{\extrarowheight}{7pt}
\small
\caption{Priors used in the model fit of the REB modulations and transit, together with the results of the fit for each of the three characterized companions.}             
\label{tab:priors}      
\centering          
\begin{tabular}{c c c c c c}     
\hline\hline       

Par.                           & Prior                     & Units          & \textbf{KOI-554}   &  \textbf{KOI-1074}     & \textbf{KOI-3728}     \\ \hline
$M_2$                          & $\mathcal{U}(0,300)$      & $M_{\rm Jup}$    & $88_{-17}^{+18}$            & $155_{-54}^{+41}$        & $83_{-13}^{+13}$             \\
$R_2 $                         & $\mathcal{U}(0,10)$       & $  R_{\rm Jup}$  & $0.77_{-0.15}^{+0.15}$      & $1.38_{-0.30}^{+0.29}$    & $2.38_{-0.51}^{+0.53}$       \\
$M_{\star}$\tablefootmark{(a)} & $\mathcal{G}(M,\sigma_M)$ & $ M_{\odot}$     & $1.13_{-0.15}^{+0.15}$      & $1.09_{-0.18}^{+0.18}$   & $2.10_{-0.30}^{+0.31}$          \\
$R_{\star}$\tablefootmark{(a)} & $\mathcal{G}(R,\sigma_R)$ & $ R_{\odot}$     & $1.08_{-0.21}^{+0.21}$      & $1.30_{-0.30}^{+0.29}$     & $4.36_{-0.85}^{+0.99}$        \\
$a $                           & $\mathcal{U}(0,0.5)$      &  au              & $0.0303_{-0.0059}^{+0.0060}$ & $0.048_{-0.010}^{+0.011}$ & $0.137_{-0.030}^{+0.032}$    \\
$e $                           & $\mathcal{U}(0,1)$        &  -            & $0.042_{-0.019}^{+0.018}$   & $0.222_{-0.079}^{+0.054}$ & $0.031_{-0.017}^{+0.016}$    \\
$\omega $                      & $\mathcal{U}(-180,180)$   &  deg.            & $159_{-30}^{+19}$           & $240_{-61}^{+14}$        & $-35_{-38}^{+48}$            \\
$i $                           & $\mathcal{U}(0,90)$       &  deg.            & $81.19_{-0.65}^{+0.54}$     & $86.16_{-0.56}^{+0.64}$  & $82.06_{-0.44}^{+0.43}$     \\
$A_B\tablefootmark{(c)} $      & $\mathcal{U}(0,5)$        &  -             & $0.14_{-0.10}^{+0.17}$     & $1.76_{-0.47}^{+0.50}$    & $0.158_{-0.094}^{+0.120}$     \\
$B\tablefootmark{(b)} $        & $\mathcal{G}(B,\sigma_B)$ &  -             & $3.71_{-0.17}^{+0.18}$      & $3.59_{-0.18}^{+0.17}$   & $3.13_{-0.13}^{+0.13}$      \\
$Sc $                          & $\mathcal{U}(-200,200)$   &  ppm             & $94_{-17}^{+18}$            & $119_{-69}^{+45}$        & $22.2_{-4.6}^{+4.1}$         \\   
\hline                  
\end{tabular}
\tablefoot{Parameters with Gaussian priors use as mean value and standard deviation those from CFOP (in the case of $M_{\star}$ and $R_{\star}$) or our calculated values and their uncertainties (B). These values can be found in Table~\ref{tab:ancillary}.
\tablefoottext{a}{We truncate the Gaussians at 0 to avoid physically meaningless solutions.}
\tablefoottext{b}{Assuming a 5,\% uncertainty.}
\tablefoottext{c}{Values above 1 account for the possibility of these targets having some intrinsic luminosity.}
}

\end{table*}